\documentclass[%
 reprint,
 amsmath,amssymb,
 aps,
]{revtex4-2}

\usepackage{graphicx}
\usepackage{dcolumn}
\usepackage{bm}
\usepackage{braket}
\usepackage{grffile}

\usepackage{tcolorbox}
\usepackage{mdframed}

\usepackage{amsmath,amssymb}
\usepackage{mathtools}
\usepackage{tikz}
\usetikzlibrary{quantikz}
\usepackage{pgfplots}
\usepackage{subcaption}

\usepackage{cprotect}
\pgfplotsset{compat=1.16}
\usetikzlibrary{positioning}
\usepackage{multirow}
\usepackage{amssymb}
\usepackage{algpseudocode}
\usepackage[boxruled, linesnumbered]{algorithm2e}
\usepackage{siunitx}
\usepackage{hyperref}

\usepackage{listings}
\usepackage{tikz}

\newcommand{\be}{\begin{equation}}
\newcommand{\ee}{\end{equation}}
\newcommand{\bea}{\begin{eqnarray}}
\newcommand{\eea}{\end{eqnarray}}
\begin{document}

\title{Response of strongly coupled fermions on classical and quantum computers}

\author{John Novak$^{1}$}
\author{Manqoba Q. Hlatshwayo$^{2}$}
\author{Elena Litvinova$^{1, 3}$}
\email{elena.litvinova@wmich.edu}
\affiliation{%
 $^1$ Department of Physics, Western Michigan University, Kalamazoo, MI, 49008, USA\\
 $^2$ National Quantum Computing Centre, Didcot, OX110QX, United Kingdom \\
 $^3$ Facility for Rare Isotope Beams, Michigan State University, East Lansing, MI, 48824, USA
}

\date{\today}
\begin{abstract}
Studying the response of quantum systems is essential for gaining deeper insights into the fundamental nature of matter and its behavior in diverse physical contexts.
Computation of nuclear response is critical for many applications, but its spectroscopically accurate description in medium-heavy nuclei in wide energy ranges remains 
particularly challenging because of the complex nature of nuclear quantum states in the high-level-density regime. 
Herein, we push the limits of configuration complexity in the classical computation of the nuclear response 
and present an algorithm with a quantum benefit for treating complex configurations. The classical computational method of approaching spectroscopic accuracy is implemented for 
medium-heavy nuclei and pioneered for the dipole response of
$^{120}$Sn, while the quantum algorithm reaching the exact solution is realized for 
the Lipkin
Hamiltonian 
to unravel the emergence of collectivity at strong coupling. 

\end{abstract} 
 
\maketitle 

Probing strongly interacting quantum systems reveals a rich interplay of fundamental interactions, quantum entanglement, many-body effects, and phase transitions. 
Modeling atomic nuclei and nuclear matter of neutron stars, in particular, relies on knowledge about nuclear compressibility, polarizability, and various decays, which is critical for many applications.
The nuclear response, the major source of this information, is especially challenging because its accurate computation requires 
thorough knowledge of nucleon-nucleon interaction in the low-energy limit of quantum chromodynamics and non-perturbative solutions for the many-body problem of strongly coupled self-bound nucleons. 

The response of interacting fermionic systems can be quantified by the equation of motion (EOM) for the propagation of a pair of fermions through the correlated medium. This equation is, in principle, exact \cite{Martin1959,Rowe1968,AdachiSchuck1989,DukelskyRoepkeSchuck1998}; however, its integral part is connected with the hierarchy of the higher-rank propagators, making the exact solution intractable for heavy nuclei. Reasonable approximations are possible to decouple the interaction kernel from this hierarchy while keeping the leading effects of emergent collectivity \cite{Martin1959,Olevano2019,LitvinovaSchuck2019}.
The formation of collective degrees of freedom appears in many nuclear structure contexts. The most evident signatures of collectivity are (i) the inability to interpret the large probabilities of transitions to some excited (collective) states in terms of the pure single-particle motion and (ii) superfluidity manifested in the nuclear rotations and spectral gaps \cite{RingSchuck1980}. 
The exact response EOM formulated in the space of the Bogoliubov quasiparticles \cite{Litvinova2022a} takes collectivity and superfluidity into account on equal footing, thus setting a rigorous microscopic framework for controllable approximations to the response of strongly interacting fermionic systems.

Early approaches to the nuclear response employed phenomenological effective forces 
and empirical couplings between the single-particle and collective modes.
Nuclear field theory (NFT) \cite{BohrMottelson1969,BohrMottelson1975,BesBrogliaDusselEtAl1976,BertschBortignonBroglia1983,Barranco2017}, quasiparticle-phonon model (QPM) \cite{Malov1976,Soloviev1992} and extensions of the Landau-Migdal theory \cite{Tselyaev1989,KamerdzhievTertychnyiTselyaev1997} were particularly advanced in the description of fragmentation mechanisms with the minimal (quasi)particle-vibration coupling ((q)PVC). With the advent of accurate nuclear energy density functionals (EDF), self-consistent qPVC approaches became available \cite{LitvinovaRingTselyaev2008,LitvinovaRingTselyaev2010,NiuNiuColoEtAl2015,Robin2019,Litvinova2023,Li2023}. 
However, although the leading-order qPVC enabled a reasonable description of the gross properties of the nuclear response, it is still insufficient for spectroscopically accurate results demanded by nuclear data, stellar astrophysics, and new physics searches at the precision frontier.

The recent effort was dedicated to the refined computation of the nuclear response. 
Bare nucleon-nucleon interactions were employed \cite{PapakonstantinouRoth2009,Knapp:2014xja,Knapp:2015wpt} and higher configuration complexities were explored \cite{Ponomarev1999,LoIudice2012,Savran2011,Tsoneva:2018ven,Lenske:2019ubp} in the studies of medium-heavy nuclei, which indicated that complex configurations are responsible for fine spectral details.  Combining the ab-initio EOM \cite{AdachiSchuck1989,DukelskyRoepkeSchuck1998,LitvinovaSchuck2019} and effective meson-exchange nucleon-nucleon interaction \cite{VretenarAfanasjevLalazissisEtAl2005,PaarNiksicVretenarEtAl2004a,PaarVretenarKhanEtAl2007}, the relativistic NFT (RNFT) introduced self-consistent computation of the response of medium-light nuclei with correlated three-particle-three-hole ($3p3h$), or six-quasiparticle ($6q$) configurations \cite{Litvinova2022a,Litvinova2023a} organized by qPVC, in the wave functions of the excited states. 
These studies revealed that (i) this configuration complexity is necessary for achieving spectroscopic accuracy, and (ii) the results may saturate quite fast with configuration complexity. We hypothesize that the latter can thus reliably serve for the uncertainty quantification of the many-body method when applied also to heavier nuclei and, further, to fermionic systems with arbitrarily strong coupling. 



In this work, we formulate an approach for the fermionic response with a variable configuration complexity and demonstrate its validity in two contexts. First, the RNFT implementation of the $6q$ configuration complexity is advanced to considerably larger model space and pioneered for medium-heavy superfluid tin nuclei in classical computation. Second, we tackle the problem with the quantum EOM (qEOM) algorithm, originally proposed for calculations of energy eigenstates of fermionic systems \cite{Ollitrault2020} and later to the single-fermion Green functions \cite{Rizzo2022}. Here, we promote the qEOM to the quantum computation of the response function and discover its quantum benefit in treating complex configurations at strong coupling for the response of the Lipkin-Meshkov-Glick (LMG) Hamiltonian \cite{Lipkin1965}. Although LMG is a generic testbed for many-body physics \cite{Delion2005,Vidal2007,Dukelsky2013} and quantum algorithms \cite{Cervia2021,Romero2022,Hlatshwayo2022,Robin2023,BeaujeaultTaudiere2024,Hlatshwayo2024}
because of its scalable coupling and integrability, its response properties were largely unknown. Our LMG study reveals the role of complex configurations in the LMG response and tracks the emergence of collectivity over the coupling regimes with quantified uncertainties.  

\begin{figure}
\begin{center}
\includegraphics[scale=0.22]{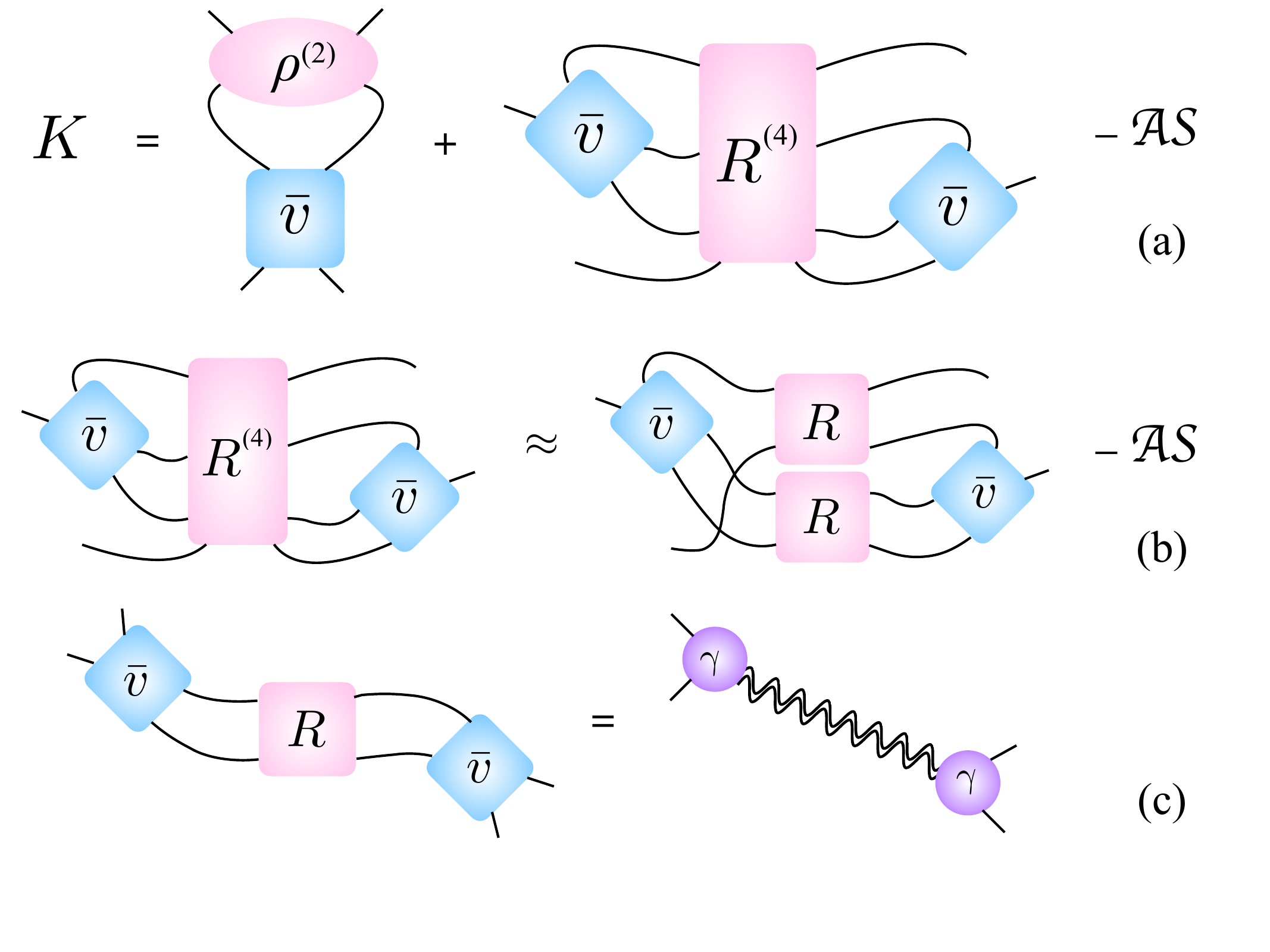}
\end{center}
\vspace{-1.1 cm}
\caption{Schematic exact interaction kernel (a), its approximation truncated at the two-body level (b), and the emergence of collective degrees of freedom (c). The symbol $\bar v$ denotes the antisymmetrized bare interaction, $\rho^{(2)}$ is the two-body density, and $R^{(4)}$ is the propagator of four quasiparticles. Connecting solid lines are assigned to quasiparticles (particles and holes), and the wiggly line denotes the phonon propagator. '$AS$' stands for omitted terms with antisymmetrized fermionic operators \cite{LitvinovaSchuck2019,Litvinova2022a}.
}
\vspace{-0.5 cm}
\label{Kernel}%
\end{figure}

In the field-theoretical formalism widely applicable across the energy scales from particle physics \cite{Popovici2010} to quantum chemistry \cite{Olevano2019}, the response of a fermionic system to a local external operator is associated with the two-time correlation function defining in-medium particle-hole (two-quasiparticle) propagation:  
\be
R_{ij,kl}(t-t') =  -i\langle T\psi^{\dagger}_i(t)\psi_j(t)\psi^{\dagger}_l(t')\psi_k(t')\rangle,
\label{phresp}
\ee
where $\braket{T...}$ stands for the chronologically ordered expectation value in the ground state and $\psi_i(t), \psi^{\dagger}_i(t)$ are the fermionic field operators in the Heisenberg picture at time $t$. The Latin subscripts stand for the single-particle basis states, and in the covariant formulation, the Hermitian adjoints are replaced by the Dirac adjoints. 
The response function contains the entire spectrum of excited states $|n\rangle$, which is commonly accessed via its  Fourier transform
\be
R_{ij,kl}(\omega) = \sum\limits_{n>0}\Bigl[ \frac{\rho^{n}_{ji}\rho^{n\ast}_{lk}}{\omega - \omega_{n} + i\delta} -  \frac{\rho^{n\ast}_{ij}\rho^{n}_{kl}}{\omega + \omega_{n} - i\delta}\Bigr],
\label{respspec}
\ee
where the poles $\omega_{n}$ 
are the excitation energies above the ground state $|0\rangle$, $\delta \to +0$, and 
$\rho^{n}_{ij} = \langle 0|\psi^{\dagger}_j\psi_i|n \rangle$ are the transition densities. The EOM for the response function is the Bethe-Salpeter-Dyson equation (BSDE), whose operator form reads:
\be
R(\omega) = R^{(0)}(\omega) + R^{(0)}(\omega)K[\rho^{(2)},R^{(4)}(\omega)]R(\omega),
\label{BSDE}
\ee
with $R^{(0)}(\omega)$ being the uncorrelated particle-hole propagator and $K(\omega)$ the complete interaction kernel. In this formulation, the kernel $K(\omega)$  plays the role of the in-medium interaction between the fermions. Its schematic form is displayed in Fig. \ref{Kernel}(a), omitting technical details available in Refs. \cite{LitvinovaSchuck2019,Litvinova2022a}. The essential feature of $K(\omega)$ is its decomposition into the static and dynamic components, where the former consists of the bare (antisymmetrized two-body) fermionic interaction $\bar v$ contracted with the correlated two-fermion density $\rho^{(2)}_{ijkl} = \langle \psi^{\dagger}_k\psi^{\dagger}_l\psi_j\psi_i\rangle$ and the latter comprises the two-time four-fermion correlation function 
$R^{(4)}_{ijkl,mnpq}(t-t') = -i\langle T({\psi^{\dagger}}_i{\psi^{\dagger}}_j\psi_k\psi_l)(t)({\psi^{\dagger}}_{q}{\psi^{\dagger}}_{p}\psi_{n}\psi_{m})(t')\rangle$ contracted with two matrix elements of $\bar v$.

The dynamic component
represents the long-range correlations associated with the system size because of the presence of the propagator $R^{(4)}(t-t')$, while the static one is practically instantaneous, having nearly the range of the bare interaction. Since $R^{(4)}$ further couples to higher-rank propagators $R^{(k)}$ with $k>4$, reasonable approximations are needed to make the EOM (\ref{BSDE}) tractable. The cluster decomposition $R^{(4)}\sim R\otimes R$ illustrated schematically in Fig. \ref{Kernel}(b) is of particular interest for nuclear systems as it enables a closed form of the BSDE with respect to the correlation function $R$ and keeps important effects of the emergent collectivity. The latter is featured in Fig. \ref{Kernel}(c) as one correlation function $R$ in the factorized kernel appearing in a double contraction with $\bar v$. The spectral expansion of $R$ (\ref{respspec}) comprises poles with factorized residues so that each contraction results in a vertex function
$\gamma^{n}_{ij} = \sum_{kl}{\bar v}_{ikjl}\rho^{n}_{lk}$. The corresponding pole $\omega_{n}$ is 
the energy of the quasibound two-quasiparticle state
(frequency of the vibrational eigenmode). 
The formed composite bosons (phonons) live in the correlated medium and play the role of mediators of the induced interaction. 
This mechanism expresses the emergence of the qPVC scale in fermionic systems with strong coupling, 
which is derivable from the underlying bare forces. In the strong coupling regime of nuclear physics, the emergent phonons are enhanced by the coherent contribution of many quasiparticle pairs so that they indeed form collective degrees of freedom, which are manifest on-shell as collective excitations. 

The giant dipole resonance (GDR) is one such mode dominating the spectra of atomic nuclei,
which makes it an ideal testbed for novel theoretical approaches. Since the exact solution for the nuclear response is beyond reach, the best benchmark for the theory is the experimental data with the given uncertainties. For this work, we have selected the recent data of Ref. \cite{Bassauer2020} for the GDR in stable tin isotopes. As the dipole spectra for these nuclei look alike, we chose to focus on the isotope $^{120}$Sn with well-expressed superfluidity in its neutron subsystem. For the static EOM kernel, we employed the leading-order meson-exchange interaction parametrized within the covariant EDF \cite{Lalazissis1997} and corrected by subtraction, restoring the consistency of the ab initio framework \cite{Tselyaev2013}, whereas the subleading orders are taken into account non-perturbatively by the dynamical kernel. 
This scheme enables entering the self-consistent cycle    
of BSDE
where the in-medium interaction requires the knowledge of multi-fermion correlation functions.
In this work, the approach to the dynamical kernel outlined in Fig. \ref{Kernel} was implemented with up to $6q$ configurations via iterating Eq. (BSDE),
so that each iteration adds two quasiparticles ($2q$) to the configuration complexity of $R(\omega)$. 
The first iteration neglecting the dynamical kernel yields $2q$ configurations, which correspond to the relativistic quasiparticle random phase approximation (RQRPA) \cite{VretenarAfanasjevLalazissisEtAl2005,PaarNiksicVretenarEtAl2004a}, dubbed relativistic EOM$^1$ (REOM$^1$), and generates the phonon characteristics so that each phonon represents a correlated $2q$ configuration. The second iteration (REOM$^2$) produces $4q$, or $2q\otimes phonon$, configurations, and the third one (REOM$^3$) explicitly resolves  $6q$, or $2q\otimes 2phonon$, configurations in the excited state wave functions. Since the number of excited states grows as the power of configuration complexity, progressively denser spectra (\ref{respspec}) are generated on each iteration. Numerical details are given in Ref. \cite{Litvinova2023a}.

\begin{figure}
\begin{center}
\includegraphics[scale=0.5]{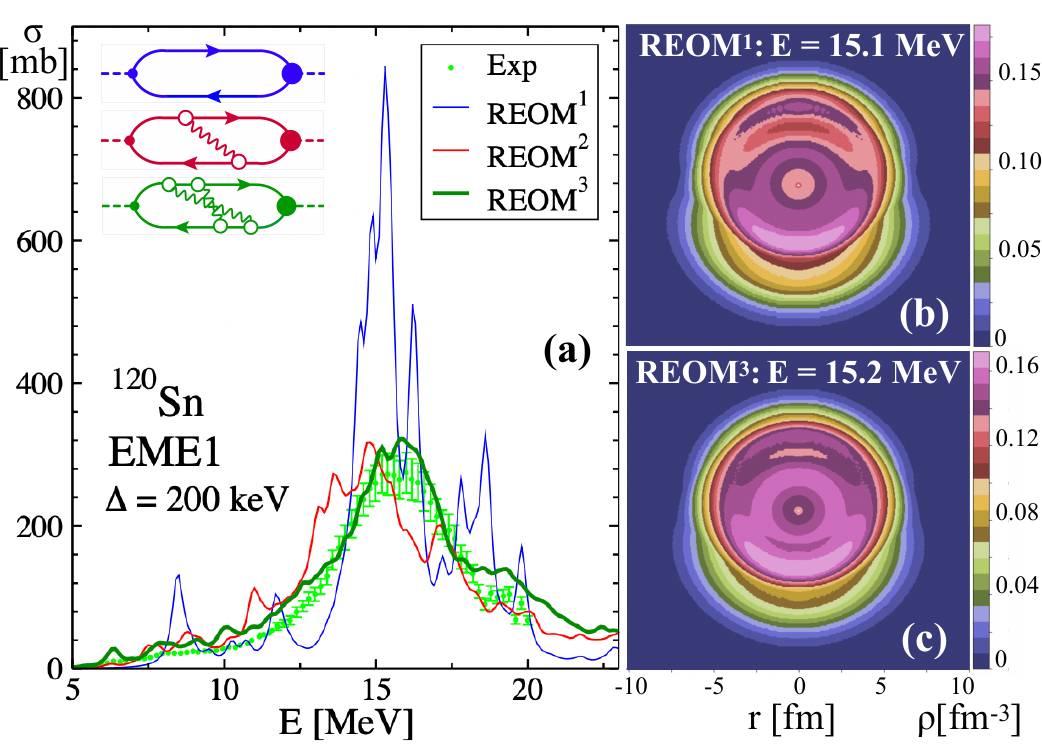}
\end{center}
\vspace{-0.5 cm}
\caption{Isovector giant dipole resonance in $^{120}$Sn in REOM$^{1-3}$ approaches identified with the major polarization diagrams in the qPVC vertex expansion, compared to data \cite{Bassauer2020} (a).  The GDR density profiles 
in REOM$^1$ (b) and REOM$^3$ (c) at centroid energies. }
\label{IVGDR}%
\vspace{-0.5 cm}
\end{figure}
The nuclear response was obtained with the dipole operator for the system of $N$ neutrons and $Z$ protons \cite{PaarVretenarKhanEtAl2007}
\bea
{ F} &=& \frac{eN}{N+Z}\sum\limits_{a=1}^Z r_aY_{1M}({{\bf r}}_a) - \frac{eZ}{N+Z}\sum\limits_{a=1}^N r_aY_{1M}({{\bf r}}_a)  
\label{opE1E2}
\eea 
defining the strength function 
$S_F(E,\Delta) = -\frac{1}{\pi}\Im\langle {F}{R}(E+i\Delta){F}^{\dagger}\rangle$,
where the imaginary part of the energy variable was taken $\Delta = 200$ keV to match the experimental resolution of the data chosen for comparison. The strength distributions in 
REOM$^{1-3}$ and the data of Ref. \cite{Bassauer2020} are displayed in Fig. \ref{IVGDR} (a). One can see that the $2q$ configurations of RQRPA are able to correctly capture the overall behavior of the GDR in terms of its centroid and total strength, while $2q\otimes phonon$ configurations included in REOM$^2$ are needed for an adequate reproduction of the spreading.  However, the REOM$^2$ strength still shows non-observed gross structures and may underestimate the centroid. The next level of configuration complexity $2q\otimes 2phonon$ achieved in REOM$^3$ enables eliminating the gross structure artifacts, corrects the centroid, and produces a smooth distribution, which confidently trends towards the experimental data.  Panels (b) and (c) illustrate the GDR's "tomography" via density profiles obtained from the residues of Eq. (\ref{respspec}), establishing the relationship between strong correlations in their dynamics and geometry. Out-of-phase oscillation of the proton and neutron Fermi liquids is evident; however, in the more complete REOM$^3$ approach, its amplitude appears smaller while density gradients are smoother in bulk, which stems from a much richer structure and mixing of states forming the spectrum (\ref{respspec}).

The method thus demonstrates that although $n \sim N$ configuration complexity is formally needed to solve the $N$-body problem exactly, in practice, even at strong coupling, a moderate complexity may be sufficient. The presented approach of cumulating it is an optimal combination of fundamentality, feasibility, and accuracy for the nuclear response. 
Once the importance of generating complex configurations is recognized while the required $n$ is not known a priori, more efficient methods can be designed for strongly coupled systems.
Among quantum algorithms, gradually becoming an attractive alternative for such systems
\cite{Dumitrescu2018,Lu2019a,Roggero2019,Roggero2020,Lacroix2020,Guzman2022,Robin2021,Robin2023,Lacroix2023}, the quantum EOM (qEOM) 
 \cite{Ollitrault2020,Asthana2023,Hlatshwayo2022,Hlatshwayo2024} is maximally aligned with this goal. Herein, we extend qEOM to the computation of the fermionic response in a hierarchy of growing-complexity approximations.

The LMG model describes a system of $N$ fermions constrained to two $N$-fold degenerate energy levels 
and interacting via 
the Hamiltonian
\bea
{H} = \epsilon {J}_z -\frac{V}{2} \left({J}_+^2 + {J}_{-}^2\right), \ \ \ \ \ \ \ \ \ \ \ \ \ \ \ \ \  \\
{J}_z = \frac{1}{2}\sum\limits_{m=1}^N\sum\limits_{\sigma=\pm } \sigma{\psi}^{\dagger}_{\sigma m}{\psi}_{\sigma m}, \ \ \
{J}_{\sigma} =\sum\limits_{m=1}^N {\psi}^{\dagger}_{\sigma m}{\psi}_{-\sigma m},
\label{LMGH}
\eea
where the index $\sigma=\pm $ differentiates the upper and lower levels and ${J}_z, {J}_{\sigma}$ satisfy the angular momentum commutation algebra.  The interaction term associated with $V$ scatters two particles from the same energy level up or down coupling the states that differ by two units of the angular momentum projection. The simplicity and scalability of the interaction allow us to eliminate the uncertainties of the nuclear forces, focusing solely on the interacting many-body problem.
The Hamiltonian (\ref{LMGH}) enables the encoding scheme limited by  $n_q = \lfloor \log{_2\left(\frac{N}{2}+1\right)} \rfloor$ qubits \cite{Hlatshwayo2022,Hlatshwayo2024} and efficiently handled by the variational quantum eigensolver (VQE) \cite{Tilly2022} finding the ground state $|0\rangle$. The hierarchy of approximations is constructed by applying the excitation operator  
\be
{O}^\dagger_n(\alpha_m) = \sum_{\alpha=1}^{\alpha_m}\sum_{\mu} \left[ X^{\alpha}_{\mu} (n) {K}^{\alpha}_{\mu} - Y^{\alpha}_{\mu} (n) \left({K}^{\alpha}_{\mu}\right)^\dagger \right],   
\label{On}
\ee
where ${K}^{1}_{\mu_{1}} = {\psi}^{\dagger}_{p}{\psi}_{h}, {K}^{2}_{\mu_{2}} = {\psi}^{\dagger}_{p}{\psi}^{\dagger}_{p'}{\psi}_{h'}{\psi}_{h}$, ... with $p(h) = \{m,+(-)\}$, generating growing-complexity excited states $|n\rangle = {O}^\dagger_n|0\rangle$ in terms of $\alpha$ particle-hole ($\alpha p\alpha h$) pairs. EOM is recast as the eigenvalue equation: 
\bea
\begin{pmatrix} \mathcal{A} &  \mathcal{B} \\ \mathcal{B}^{*} & \mathcal{A}^{*} \end{pmatrix}
\begin{pmatrix} X(n) \\ Y(n) \end{pmatrix} = \omega_{n}
\begin{pmatrix} \mathcal{C} & \mathcal{D} \\ -\mathcal{D}^{*} & -\mathcal{C}^{*} \end{pmatrix}
\begin{pmatrix} X(n) \\ Y(n) \end{pmatrix},
%
\label{GEE}
\eea
with $\mathcal{A}_{\mu\nu}^{(\alpha\beta)} = \braket{ \left[ {K}^{\alpha\dagger}_{\mu}\left[{H},{K}^{\beta}_{\nu} \right]  \right] }$, 	
$\mathcal{C}_{\mu\nu}^{(\alpha\beta)} = \braket{ \left[{K}^{\alpha\dagger}_{\mu},{K}^{\beta}_{\nu}  \right]}$,  
$\mathcal{B}_{\mu\nu}^{(\alpha\beta)} = - \braket{ \left[{K}^{\alpha\dagger}_{\mu},\left[{H},{K}^{\beta\dagger}_{\nu} \right]  \right] }$, and  
$\mathcal{D}_{\mu\nu}^{(\alpha\beta)} = - \braket{ \left[ {K}^{\alpha\dagger}_{\mu},{K}^{\beta\dagger }_{\nu} \right] }$.
Eq. (\ref{GEE}) is post-processed on a classical computer, for which powerful methods exist \cite{Nakatsukasa2007,Bjelcic2020,Litvinova2022a}. 
\begin{figure}
\begin{center}
\includegraphics[scale=0.3]{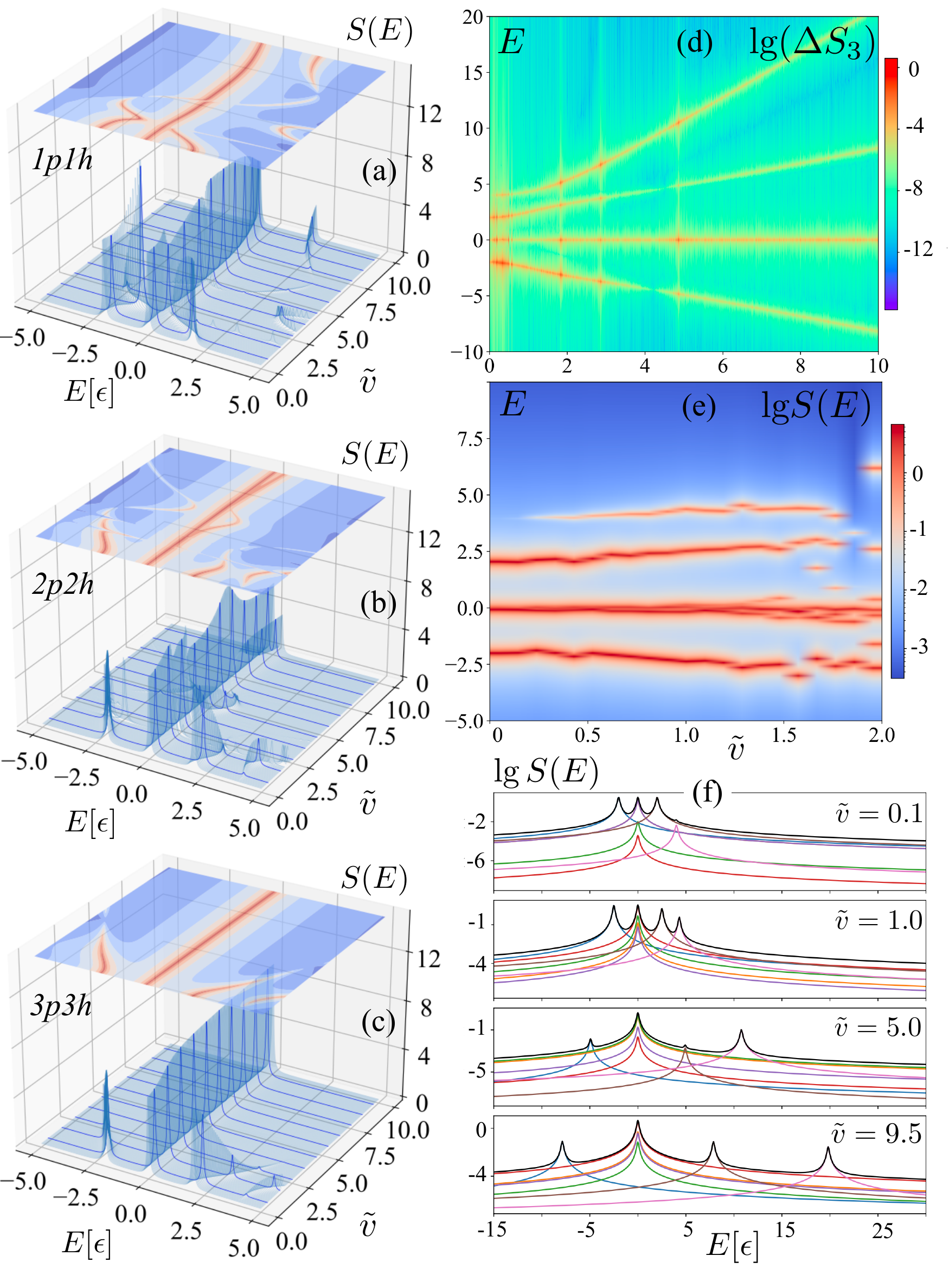}
\end{center}
\vspace{-0.5 cm}
\caption{Strength distributions for the LMG ($N=8$) model obtained on a digital simulator for increasing configuration complexity (a)-(c); deviation of the $3p3h$ strength functions from the exact ones (d); $3p3h$ strength obtained on the IBM Quantum computers (e); $3p3h$ (exact) strength function profiles (f).}
\vspace{-0.5 cm}
\label{LMG}%
\end{figure}
The major advantage of the qEOM algorithm is the efficient evaluation of the ground-state expectation values for the matrix elements $\mathcal{A},\mathcal{B},\mathcal{C},\mathcal{D}$ \cite{Ollitrault2020}.  
They are expressed by products of Pauli gates $P_k \in \{ \mathbb{I}, X, Y, Z\}$:
 \be
\mathcal{A} = \sum_{i} a_{i} \braket{ \{P_0 \otimes P_1 \otimes \ldots P_{n_q}\}^i}
\label{Pauli}
\ee
and analogous decompositions for $\mathcal{B}$, $\mathcal{C}$, and $\mathcal{D}$.
The number of quantum measurements $\braket{ P_0 \otimes P_1 \otimes \ldots P_{n_q}}$ is limited by $n_q$.  Increasing the configuration complexity $\alpha$ needed for strong coupling increases the number of terms in the sum (\ref{Pauli}) but does not significantly affect the number of measurements. This is a spectacular quantum benefit, which further extends to the strength function. For a trial excitation operator $F = \sum\limits_{ph} (K^1_{ph} + K^{1\dagger}_{ph})$, it reads
\bea
S_F(E;\Delta) &=& \frac{\Delta}{\pi}\sum\limits_{n} \frac{|\bra{0}{F}\ket{n}|^2}{(E-\omega_n)^2 + \Delta^2},  
\label{TrME}
\\
\bra{0}{F}\ket{n} &=& 
\sum_{{ph},\alpha\mu}\Bigl( X^{\alpha}_{\mu} (n) {\mathcal M}_{ph,\mu} 
+ Y^{\alpha}_{\mu} (n)  {\mathcal M}^{\ast}_{ph,\mu} 
\Bigr) \nonumber
\eea
with ${\mathcal M}_{ph,\mu} = {\mathcal C}_{ph,\mu} + {\mathcal D}^{\ast}_{ph,\mu}$, being completely determined by the solutions of Eq. (\ref{GEE}) and its matrix elements already measured on a quantum computer. 

The strength functions for the $N = 8$ LMG model smoothed by $\Delta = 0.05\epsilon$ are displayed in Fig. \ref{LMG}. Panels (a)-(c) show the strength as a function of energy $E/\epsilon$ and interaction  $\tilde v = (N-1)V/\epsilon$ for the even energy levels obtained in $\alpha_m = \{1,2,3\}$ approximations on a simulator. The two-dimensional heatmaps are shown on top of the panels to guide the eye. The even part of the Hamiltonian is mapped on three qubits constraining $\alpha_m = 3$ as the maximal configuration complexity. The increase of  $\alpha_m$ gradually improves the description heading towards a clean LMG energy spectrum at $\alpha_m = 3$, which accurately reproduces the exact energies \cite{Hlatshwayo2024}. The same trend is seen in the strength functions, which at $\alpha_m < 3$ exhibit artificial peak structures causing decoherence at strong coupling and at $\alpha_m = 3$ reproduce the exact solution. The latter is illustrated in panel (c): both the $\alpha_m = 3$ and exact solutions trend toward the strength concentrated in a single peak at large $\tilde v$. This trend holds up to at least
$\tilde v \sim 100$, and all the approaches satisfy the energy-weighted sum rule. Panel (d) illustrates the accuracy of the $\alpha_m = 3$ approach in terms of deviation from the exact strength functions: the median error and 90th percentile are 1.4$\cdot 10^{-9}$ and 3.5$\cdot 10^{-7}$, respectively. Hardware computation illustrated in panel (e) was performed on IBM Quantum devices using the optimized circuit \cite{Hlatshwayo2024} up to $\tilde v = 2$ and also showed the best quality of description at $\alpha_m = 3$.  

At $\tilde v = 0$, all three approaches generate essentially identical spectra, which consist of three equidistant and equal-height peaks. With the increment of $\tilde v$, the 
levels repel, and the strength redistributes: the peak at $E=0$ gradually grows while the others are suppressed. The inaccuracies of the  $\alpha_m = \{1,2\}$ approximations also increase with $\tilde v$, whereas the $\alpha_m = 3$ calculation demonstrates a clear picture of strength collectivization into a single dominant peak. In this way, we trace the emergence of collectivity via its gradual accumulation with the strengthening of interaction. Panel (f) of Fig. \ref{LMG} shows the $S(E)$ profiles obtained from the 
$\alpha_m = 3$ simulation for each energy solution represented by separate curves, further clarifying the strength formation across the coupling regimes. One can distinguish, in particular, the two mechanisms of emerging collectivity: (i) the accumulation of the strength by a single transition at $E = 0$ due to its growing amplitude and (ii) the coherent contribution of multiple transitions. 

In summary, we presented a microscopic approach to the response of strongly coupled fermionic systems with variable complexity of their quantum states. We showed that (i) the degree of complexity is a reliable classifier of the accuracy of the many-body approach with an established link to the exact theory, and (ii) accounting for emergent collectivity plays a critical role in obtaining spectroscopically accurate results. 
%
Our method demonstrated that many-body correlations of growing complexity are efficiently generated if 
the emergent collectivity is utilized to order complex configurations in large ($N\approx120$) systems.
In particular, the GDR in $^{120}$Sn is reproduced with nearly spectroscopic accuracy with correlated $2q\otimes 2phonon$ configurations. We argued that for nuclear spectral calculations, an efficient method of tackling configurations with varying complexity is the key ingredient, and such a method is accessible via quantum algorithms. The quantum EOM is implemented for the response of the LMG system, revealing a quantum benefit when building a hierarchy of approximations ordered by configuration complexity $\alpha_m$ due to the independence of the number of required quantum measurements on this parameter. 
Calculations on a digital simulator demonstrated the capability of qEOM to gradually approach the exact solution with growing $\alpha_m$, efficiently accessing emergent collectivity via entanglement, which holds up to arbitrarily strong coupling. The available IBM Quantum hardware enables reliable computation up to the effective coupling strength of 1.0-1.5, which is in the range of nuclear structure physics. The method shows a reasonable scaling with the particle number, opening a new avenue to studies of nuclear and other strongly coupled many-body systems on quantum computers.




We acknowledge funding from US-NSF through the Grant PHY-2209376 and CAREER Grant PHY-1654379.
We appreciate cloud access to IBM Quantum computers to run hardware calculations for this work. 


\bibliography{BibliographyDec2023}

\end{document}